\documentclass[twocolumn,times]{aastex62}
\usepackage{rotating}
\usepackage{graphicx}
\usepackage{appendix}
\usepackage{float}
\usepackage{hyperref}

\definecolor{forestgreen}{rgb}{0.13, 0.55, 0.13} %forestgreen5web)

\newcommand{\angstrom}{\mbox{\normalfont\AA}}

\newcommand{\BE}{\begin{equation}}
\newcommand{\EE}{\end{equation}}
\newcommand{\BA}{\begin{align}}
\newcommand{\EA}{\end{align}}
 \renewcommand{\fig}[1]{Figure~\ref{fig_#1}}
 
 \newcommand{\figs}[2]{Figures~\ref{fig_#1} and \ref{fig_#2}}
 
 \newcommand{\sect}[1]{Section~\ref{sect_#1}}

 \newcommand{\tab}[1]{Table~\ref{tab_#1}}

\graphicspath{{./}{figures/}}

\received{XXXX, 2019}
\revised{XXXX, 2019}
\accepted{\today}
\submitjournal{ApJ}

\shorttitle{Plasma Composition in AR12673}
\shortauthors{Baker et al.}
\begin{document}

\title{Can Sub-photospheric Magnetic Reconnection Change the Elemental Composition in the Solar Corona?}

\author[0000-0002-0665-2355]{Deborah Baker}
\affil{University College London, Mullard Space Science Laboratory, Holmbury St. Mary, Dorking, Surrey, RH5 6NT, UK}

\author[0000-0002-2943-5978]{Lidia van Driel-Gesztelyi}
\affiliation{University College London, Mullard Space Science Laboratory, Holmbury St. Mary, Dorking, Surrey, RH5 6NT, UK}
\affiliation{LESIA, Observatoire de Paris, Universit\'e PSL, CNRS, Sorbonne Universit\'e, Univ. Paris Diderot, Sorbonne Paris Cit\'e, 5 place Jules Janssen, 92195 Meudon, France}
\affiliation{Konkoly Observatory, Research Centre for Astronomy and Earth Sciences, Hungarian Academy of Sciences, Konkoly Thege \'ut 15-17., H-1121, Budapest, Hungary}

\author[0000-0002-2189-9313]{David H. Brooks}
\affiliation{College of Science, George Mason University, 4400 University Drive, Fairfax, VA 22030, USA}

\author[0000-0001-8215-6532]{Pascal D\'emoulin}
\affiliation{LESIA, Observatoire de Paris, Universit\'e PSL, CNRS, Sorbonne Universit\'e, Univ. Paris Diderot, Sorbonne Paris Cit\'e, 5 place Jules Janssen, 92195 Me udon, France}

\author[0000-0001-7809-0067]{Gherardo Valori}
\affiliation{University College London, Mullard Space Science Laboratory, Holmbury St. Mary, Dorking, Surrey, RH5 6NT, UK}

\author[0000-0003-3137-0277]{David M. Long}
\affiliation{University College London, Mullard Space Science Laboratory, Holmbury St. Mary, Dorking, Surrey, RH5 6NT, UK}

\author[0000-0002-3362-7040]{J. Martin Laming}
\affiliation{Space Science Division, Naval Research Laboratory, Code 7684, Washington, DC 20375, USA}

\author[0000-0003-0774-9084]{Andy S. H. To}
\affiliation{University College London, Mullard Space Science Laboratory, Holmbury St. Mary, Dorking, Surrey, RH5 6NT, UK}

\author[0000-0001-7927-9291]{Alexander W. James}
\affiliation{European Space Astronomy Centre, Urb. Villafranca del Castillo, E-28692 Villanueva de la Cañada, Madrid, Spain}

\begin{abstract}

Within the coronae of stars, abundances of those elements with low first ionization potential (FIP) often differ from their photospheric values. 
The coronae of the Sun and solar-type stars mostly show enhancements of low-FIP elements (the FIP effect) while more active stars such as M-dwarfs have coronae generally characterized by the inverse-FIP effect (I-FIP).
Here we observe patches of I-FIP effect solar plasma in AR 12673, a highly complex $\beta$$\gamma$$\delta$ active region.
We argue that the umbrae of coalescing sunspots and more specifically strong light bridges within the umbrae, are preferential locations for observing I-FIP effect  plasma.
Furthermore, the magnetic complexity of the active region and major episodes of fast flux emergence also lead to repetitive and intense flares. 
The induced evaporation of the chromospheric plasma in flare ribbons crossing umbrae enables the observation of four localized patches of I-FIP effect  plasma in the corona of AR 12673.
These observations can be interpreted in the context of the ponderomotive force fractionation model which predicts that plasma with I-FIP effect composition is created by the refraction of waves coming from below the chromosphere.
We propose that the waves generating the I-FIP effect plasma in solar active regions are generated by sub-photospheric reconnection of coalescing flux systems.
Although we only glimpse signatures of I-FIP effect fractionation produced by this interaction in patches on the Sun, on highly active M-stars it may be the dominant process.
\end{abstract}


%%%%%%%%%%%%%%%%%%%%%%%%%%%%%%%%%%%%%%%%%%%%%%%%%%%%%%%%%%%%%%%%%%%%%%%%%%%%%%%%%%
\section{Introduction}
\label{sect_intro}
Recent spectroscopic observations of the Sun have shown how the elemental composition of the solar atmosphere varies in space and time.
The variability of composition appears to be intrinsically linked to the distribution and evolution of the Sun's magnetic field on all scales \citep[e.g.][]{brooks17,baker18}.
More generally, the overall composition of a star's corona depends on the first ionization potential (FIP) of the main elements comprising the corona.
In solar-type stars, low-FIP elements are observed to be over-abundant relative to their photospheric abundances, while high-FIP elements maintain their photospheric abundances \citep[the FIP effect; e.g.][]{laming95,wood10,testa15,laming15,brooks17}. 
In cooler, more active stars such as M-dwarfs, low-FIP/high-FIP elements are under-/over-abundant in the corona compared to the photosphere \citep[the inverse FIP or I-FIP effect;][]{laming15,brooks18}.  
\citet{wood10} and \citet{wood18} first established an almost linear relationship between stellar composition and spectral type F to M in the X-ray spectra of moderately active stars (X-ray luminosity $<$10$^{29}$ ergs s$^{-1}$).
In addition, a dependence has been found of coronal composition on stellar magnetic activity, with high-activity stars having I-FIP-effect coronae and low-activity stars like our Sun having FIP effect dominated coronae \citep[][]{audard03,g-a09,testa15}. 

One plasma fractionation model that is able to account for the FIP and I-FIP effects observed in stars is the ponderomotive force model \citep{laming15}. 
The model invokes the ponderomotive force exerted by Alfv\'en waves giving rise to ion--neutral separation in the chromosphere of the Sun and other stars. 
Fractionation takes place in the chromosphere where the temperature and density gradients are high and temperatures are such that low-FIP elements are mainly ionized and high-FIP elements remain neutral.
The direction of the ponderomotive force determines whether low-FIP elements become enhanced or depleted in the corona.
Downward propagating Alfv\'en waves generated by magnetic reconnection in the corona are mostly reflected back into the corona at the high density gradient in the chromosphere, creating an upward-directed ponderomotive force acting on the ions \citep{laming17}. 
When the plasma is observed in the corona, the low-FIP elements are over-abundant relative to the photosphere.
An inverse FIP effect is created when upward-traveling waves undergoing reflection or refraction back downwards produce a downward-directed ponderomotive force acting on the ions in the chromosphere. 
Such upward-propagating waves can be generated, for instance, by magneto-acoustic waves originating from below the chromosphere that are mode converted to fast mode waves at the plasma $\beta$ = 1 boundary.
As a consequence of the downward ponderomotive force, the plasma is then depleted of low-FIP elements \citep{laming15,brooks18}.
In this article, we present observations that suggest a more specific origin for the waves producing the I-FIP  effect, namely a sub-photospheric one.

Though the Sun's corona is dominated by the FIP effect \citep{laming95,brooks15,brooks17}, recent spectral scans taken by the EUV Imaging Spectrometer (EIS) \citep{culhane07} on board the \emph{Hinode} spacecraft \citep{kosugi07} have shown highly localized regions of I-FIP effect plasma near sunspots in flare spectra \citep{doschek15,doschek16,doschek17}. 
\citet{doschek17} proposed that the FIP effect `shuts down' near sunspots where p-modes are suppressed, leading to areas of photospheric, weak I-FIP, or I-FIP effect plasma, while the rest of the active region has coronal composition. 

At low spatial resolution, solar coronal abundances in flares have been previously analyzed and shown to be very different from quiescent conditions \cite[e.g.][]{sylwester84,feldman90,warren14}, having a tendency to be close to photospheric values. 
Composition studies of stellar flares provide strong indications of a similar effect, i.e. that the elemental abundances tend to approach photospheric values both on FIP-effect and I-FIP-effect dominated stars \citep[e.g.][]{nordon08,laming09}.
Recently, \cite{katsuda20} reported the I-FIP effect in four large X-class flares (X17.0, X5.4, X6.2 from AR 10808 and X9.0 from AR 10930) derived from Earth albedo X-ray spectra from the imaging spectrometer on board the \emph{Suzaku} astronomical satellite.

\citet{baker19} analyzed in detail  \emph{Hinode}/EIS observations of  plasma composition during the decay phase of an M-class flare in AR 11429.
Patches of I-FIP effect  plasma appeared in a highly sheared emerging flux region above sunspot umbrae 10 min after the flare peak and disappeared 40 min later.
The authors proposed that  sub-chromospheric reconnection of highly sheared coalescing strands of the same polarity magnetic field played a key role in the creation of I-FIP effect  plasma within the highly complex magnetic field of the active region. 
During episodes of strong emergence, flux approached and interacted with pre-existing magnetic field, forcing the coalescence of the smaller flux fragments into growing, coherent umbrae surrounded by common penumbrae in two of the sunspots.
The convergence of magnetic field in the location of the coalescing umbrae is highly suggestive of the presence of sub-chromospheric magnetic reconnection.
Reconnection that occurs below the region of plasma fractionation in the chromosphere generates a fast-mode wave flux in the direction required by the ponderomotive force fractionation model to produce I-FIP effect  plasma \citep{laming15}.
The fractionated plasma is then evaporated into the corona by flaring and observed by \emph{Hinode}/EIS when flare ribbons cross the coalescing umbra.

In this paper, magnetic field and continuum observations of the most unusual and complex magnetic field of AR 12673 are combined with \emph{Hinode}/EIS scans to investigate the presence of I-FIP effect  plasma.
We confirm the findings of \citet{baker19} and provide evidence in support of the generation of I-FIP effect plasma by sub-photospheric reconnection in coalescing sunspot umbrae with strong light bridges (LBs).
The paper is organized as follows: \sect{b_obs} describes the magnetic field evolution, \sect{eis_obs} details the \emph{Hinode}/EIS observations, \sect{interp} provides our analysis and interpretation of the observations of anomalous plasma composition in the context of the ponderomotive force fractionation model, and in \sect{conc} we present our conclusions.

%%%%%%%%%%%%%%%%%%%%%%%%%%%%%%%%%%%%%%%%%%%%%%%%%%%%%%%%%%%%%%%%%%%%%%%%%%%%%%%%%%
\section{Magnetic Field Evolution of AR 12673}
\label{sect_b_obs}
AR 12673 was visible on the Sun from 2017 August 28 to September 10.
Its magnetic field evolved into one of the most complex structures observed during solar cycle 24 \citep{lvdg19}.
The sunspot group's Mount Wilson magnetic classification was $\beta$$\gamma$$\delta$, the total unsigned flux exceeded 5$\times$10$^{22}$ Mx and the projected whole spot area peaked on September 6 at $\sim$2,100 MSH (millionths of a solar hemisphere; from the Debrecen Photoheliographic Database: \url{http://fenyi.solarobs.csfk.mta.hu/DPD/2017/index.html}). 
Overall activity included 4 X-, 27 M-, and 55 C-class flares, making AR 12673 the most flare productive of solar cycle 24.
Several significant coronal mass ejections (CMEs) were launched during its disk passage \citep{redmon18}.
Furthermore, this `monster' active region exhibited highly unusual characteristics and extreme behavior including the fastest flux emergence ever recorded \citep{sun18}, the strongest transverse \citep{wang18} and coronal magnetic fields \citep{anfinogentov19}.

\begin{figure*}[htp]
\centering
\includegraphics[width=0.495\textwidth]{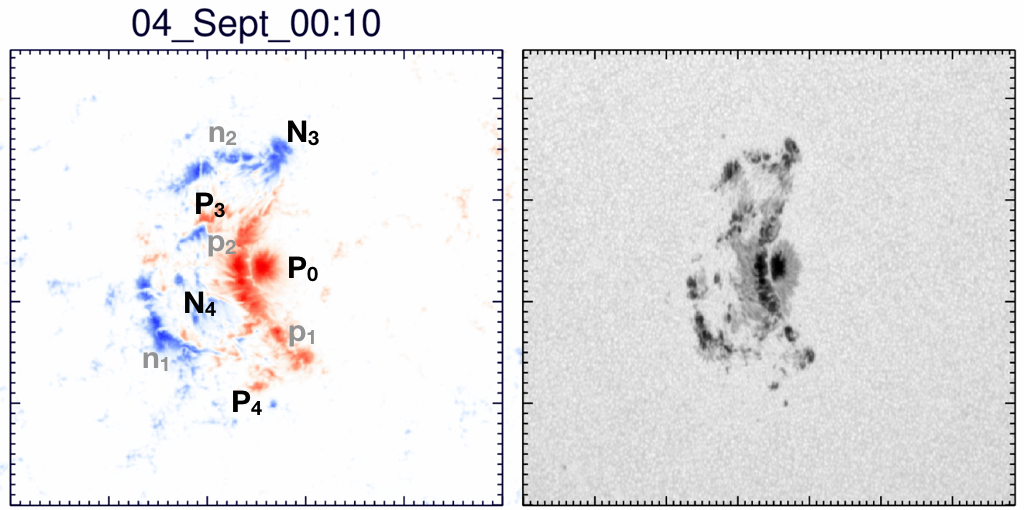}
\includegraphics[width=0.495\textwidth]{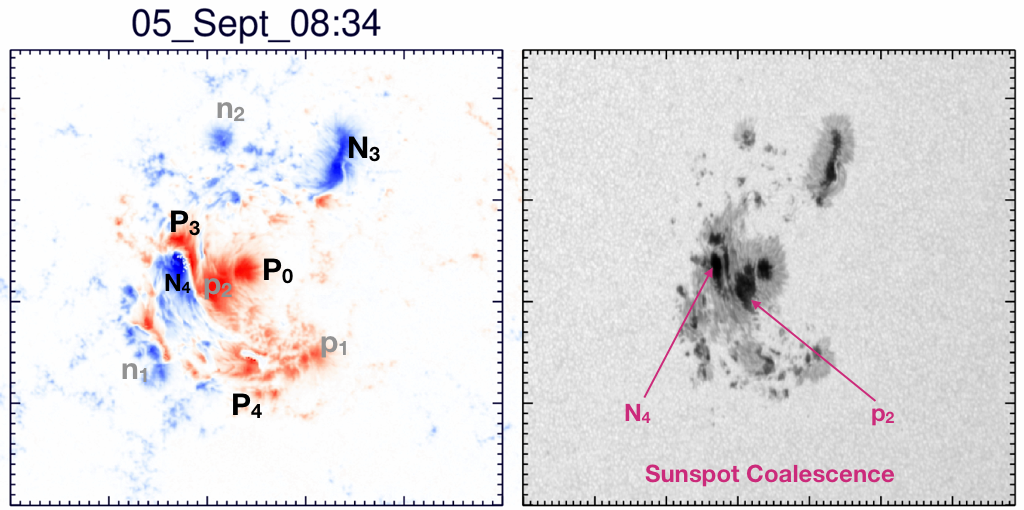}
\includegraphics[width=0.495\textwidth]{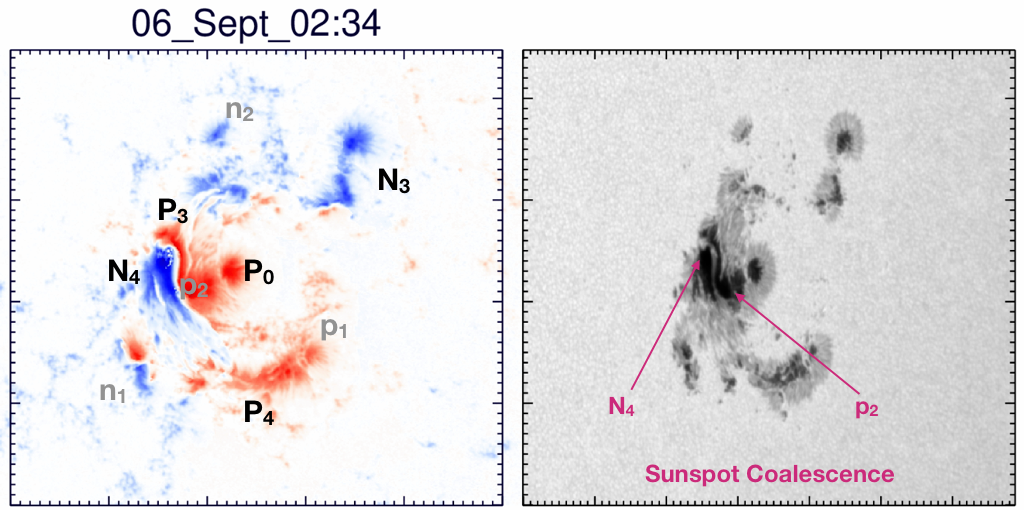}
\includegraphics[width=0.495\textwidth]{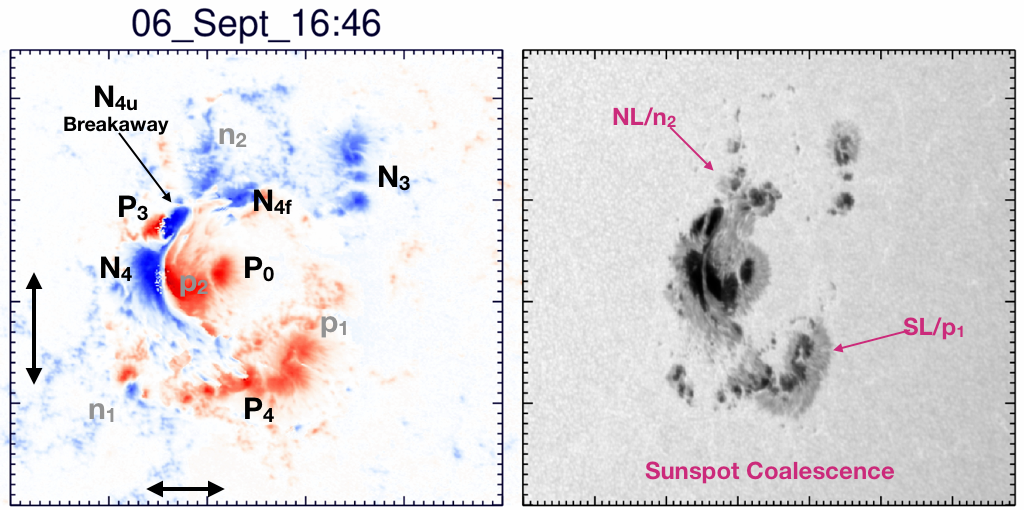}
\includegraphics[width=0.495\textwidth]{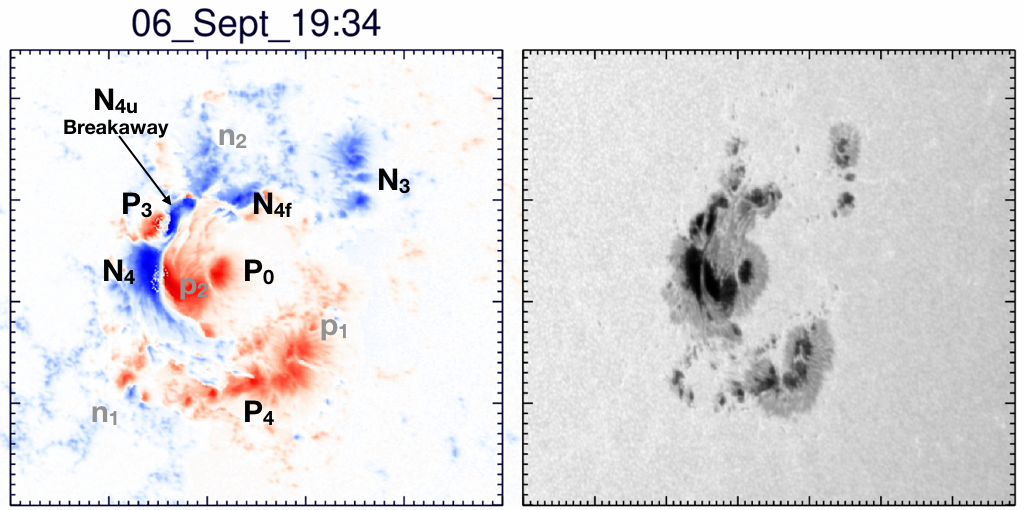}
\includegraphics[width=0.495\textwidth]{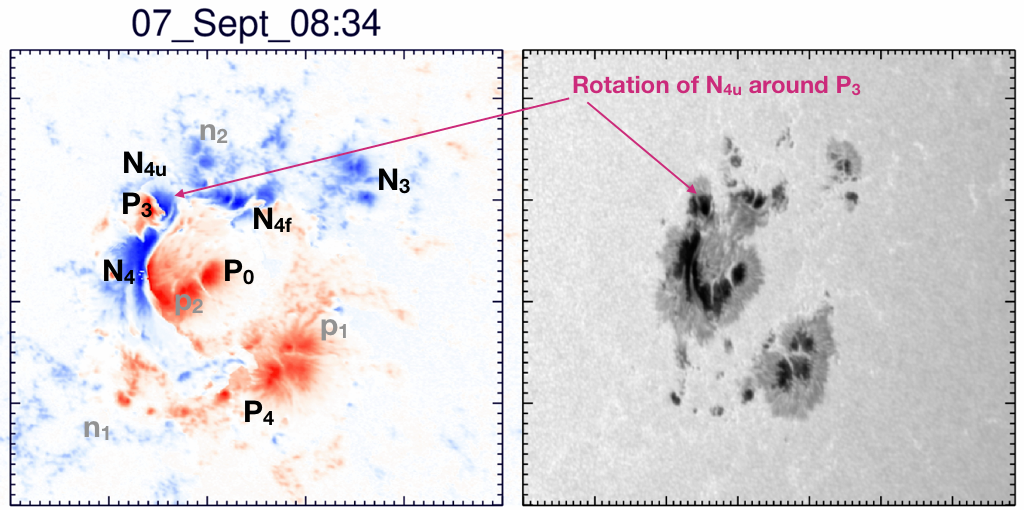}
\caption{ \emph{SDO}/HMI SHARP radial magnetic field, B$_{r}$, and 6173 $\angstrom$ continuum images \citep{bobra14} during the evolution in AR 12673.  Middle and lower panels correspond to the times of when \emph{Hinode}/EIS observed I-FIP effect  plasma (Figures \ref{fig_study559}, \ref{fig_eisobs1}, \ref{fig_eis_sep7}).  
Major components of the flux systems are labeled in each image: arcade field n$_{1}$--p$_{1}$ and n$_{2}$--p$_{2}$ (labels in gray); main field N$_{3}$--P$_{3}$ and N$_{4}$--P$_{4}$ (labels in black), and pre-existing, well anchored spot P$_{0}$. 
The U-loop is formed in between N$_{4}$--P$_{3}$ as well as N$_{4u}$--P$_{3}$ (bottom panels).  
Near the end of the emergence phase, polarities N$_{4u}$ and N$_{4f}$ detached from the north part of N$_{4}$.  
Note that n and N notations are used for negative, p and P notations for positive polarity umbrae. 
Red/blue corresponds to positive/negative B$_{r}$ in the magnetograms.
See the included movie Fig1$\_$movie.mp4 of this figure.  
For ease of comparison between figures, black arrows are used to show the distances from the center of the major polarities.  At 16:46 UT on September 6, the vertical arrow is $\sim$50$\arcsec$ in length between the centers of P$_{0}$ and P$_{4}$.  The horizontal arrow is $\sim$30$\arcsec$ in length from the centers of P$_{0}$ and N$_{4}$.  The center of P$_{0}$ is X = 590$\arcsec$, Y = -242$\arcsec$.  This convention is repeated for all figures with maps. 
\label{fig_br_map}}
\end{figure*}

\fig{br_map} shows selected co-temporal radial magnetic field and continuum images from the Helioseismic and Magnetic Imager \citep[HMI;][]{scherrer12} on board the \emph{Solar Dynamics Observatory} \citep[\emph{SDO};][]{scherrer12}.
The data are from the Spaceweather HMI Active Region Patch \citep[SHARP;][]{bobra14} pipeline.
The enclosed movie of the figure is entitled Fig1$\_$movie.mp4.  
A full account of the structure and evolution of the magnetic field leading to the first direct evidence of sub-photospheric reconnection is provided in \citet{lvdg19}. 
The main features are summarized here using the same notation.

%======================Figure 1 ============================
%======================Figure 1 ============================

AR 12673 was formed from the emergence and subsequent interaction of three flux systems labeled in \fig{br_map}: a long-lived, stable positive spot P$_{0}$; an arcade system formed of n$_{1}$--p$_{1}$ and n$_{2}$--p$_{2}$; and the main flux system composed of N$_{3}$--P$_{3}$ and N$_{4}$--P$_{4}$ (the arcade/main flux systems are labeled in gray/black in Figure \ref{fig_br_map}).
Early on September 2, bipole n$_{1}$--p$_{1}$ emerged to the south-east of pre-existing P$_{0}$.
Bipole n$_{2}$--p$_{2}$ followed one day later and emerged to the north-east of P$_{0}$.
Very quickly the two bipoles formed a coherent arcade system with a C-shaped magnetic polarity inversion line (PIL), as the fast moving emerging flux wrapped around the well-anchored P$_{0}$.
On September 3, the main field comprising bipoles N$_{3}$--P$_{3}$ and N$_{4}$--P$_{4}$ emerged in a dominantly N--S direction underneath the E--W oriented arcade system represented by n$_{1}$--p$_{1}$ and n$_{2}$--p$_{2}$.
At the location where the main system emerged into the arcade system lying above it, an M-shaped flux tube was present.  It was formed by a concave-up U-loop \citep{lvdg00} between N$_{4}$ and P$_{3}$ and two lobes of concave-down $\Omega$ loops, P$_{3}$--N$_{3}$ in the north and P$_{4}$--N$_{4}$ in the south \citep{fan03,toriumi14}. 

Key aspects of the active region's magnetic field evolution occurred in the time period of September 4--6 leading up to the times of the X2.2 and X9.3 flares.
First, during the ongoing emergence of the constituent bipoles of the three flux systems, strands of same polarity magnetic field coalesced to form coherent sunspot umbrae in the active region core to the east of P$_{0}$ (at p$_{2}$ and N$_{4}$) and at its periphery to the north and south-west at n$_{2}$ and p$_{1}$, respectively.
These were also the locations where significant LBs formed between merging umbral flux strands.
Coalescence was driven by especially strong and fast episodes of flux emergence on September 4--6 (see Fig1$\_$movie.mp4).
Second, early on September 6, the north-western part of N$_{4}$ detached and began moving northward. 
This detached polarity penetrated the `wall' of opposite polarity field along P$_{3}$--p$_{2}$ (\fig{br_map}, middle panels), then it broke in two pieces.  
One piece, N$_{4u}$, mostly rotated counter-clockwise around P$_{3}$ indicating that it was still bound by the U-loop anchorage (\fig{br_map}, bottom panels).  
The other piece, N$_{4f}$, moved rapidly northward, then westward in the direction of N$_{3}$. 
 On September 7, the main flux system --at least temporarily-- ceased its emergence into the E--W arcade flux system thereby removing the major driver of the unusual and extreme magnetic field evolution in AR 12673.
 
%======================Figure 2 ============================
\begin{figure}[tb]
\centering
\includegraphics[width=0.48\textwidth]{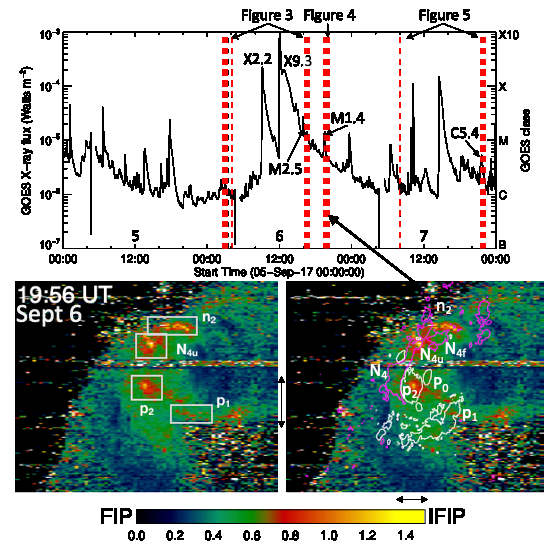}
\caption{Top panel:  GOES 1--8 $\angstrom$~soft X-ray light curve from 00:00 UT on 2017 September 5 to 00:00 UT on 2017 September 8.  Red dashed lines indicate the times of \emph{Hinode}/EIS rasters used in this study.  The thicker lines represent a series of nearby observations.
   Bottom panels:  \emph{Hinode}/EIS Ar {\sc xiv}/Ca {\sc xiv} ratio map at 19:56 UT on September 6 overplotted with boxes defining the regions used in \tab{regions} and throughout the text (left) and \emph{SDO}/HMI contours of $\pm$ 500 G of white/magenta for positive/negative polarities (right).
  The color bar scale shows the FIP effect as blue/green, photospheric composition as orange, I-FIP effect as yellow.  (P$_{0}$ is located at X = 611$\arcsec$, Y = -241$\arcsec$ at 19:56 UT).\label{fig_goes}}
\end{figure}
%======================Figure 2 ============================

%%%%%%%%%%%%%%%%%%%%%%%%%%%%%%%%%%%%%%%%%%%%%%%%%%%%%%%%%%%%%%%%%%%%%%%%%%%%%%%%%%
\section{Observations of AR 12673}\label{sect_eis_obs}
In this analysis, we focus on the three day period encompassing the largest flare of solar cycle 24, the X9.3 flare that peaked at $\sim$11:53 UT on September 6.
\fig{goes} shows the \emph{GOES} 1--8 $\angstrom$ soft X-ray light curve from 00:00 UT on September 5 -- 00:00 UT on September 8 when \emph{Hinode}/EIS was observing AR 12673 employing a multitude of studies and scanning modes.
However, not all studies contain suitable emission lines from high- and low-FIP elements for measuring plasma composition.
The basic details of the studies used here are provided in \tab{study}.  
The times of the \emph{Hinode}/EIS~composition observations are indicated by the dashed red lines plotted with the \emph{GOES} soft X-ray curve in \fig{goes}.

%======================Table 1============================
\begin{table}[t]
	\centering
	\caption{\emph{Hinode}/EIS study details.}
	\label{tab_study}
	\begin{tabular}{ll} % four columns, alignment for each
		\hline
		Study Name & HIC2$\_$SCAN$\_$201$\times$512\\
		Composition  & Ca {\sc xiv} 193.87 $\angstrom$ and Ar {\sc xiv} 194.40 $\angstrom$\\
       	Field of view & 210$\arcsec$ $\times$ 512$\arcsec$\\
		Rastering & 2$\arcsec$ slit, 21 positions, 10$\arcsec$ coarse steps \\
        Exposure Time &15 s\\
	\hline
	    Study name & Atlas$\_$60\\
		Composition Lines & Full spectral atlas including\\
		& Ca {\sc xiv} 193.87 $\angstrom$ and Ar {\sc xiv} 194.40 $\angstrom$\\
       	Field of view & 120$\arcsec$ $\times$ 160$\arcsec$\\
		Rastering & 2$\arcsec$ slit, 60 positions, 2$\arcsec$ steps \\
        Exposure Time &60 s\\
    \hline
		Study name & FlareResponse01\\
		Composition Lines &  Ca {\sc xiv} 193.87 $\angstrom$ and Ar {\sc xiv} 194.40 $\angstrom$\\
       	Field of view & 240$\arcsec$ $\times$ 304$\arcsec$\\
		Rastering & 2$\arcsec$ slit, 80 positions, 3$\arcsec$ coarse steps \\
        Exposure Time &5 s\\
		\hline
			\end{tabular}
\end{table}
%======================Table 1============================

%======================Figure 3 ============================
\begin{figure*}
\centering
\includegraphics[width=0.6\textwidth]{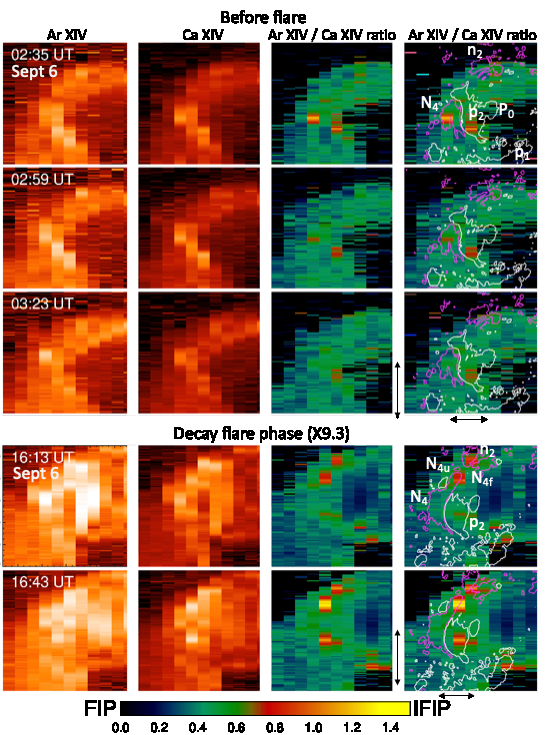}
\caption{Left to right:  \emph{Hinode}/EIS Ar {\sc xiv} 194.40 $\angstrom$~and Ca {\sc xiv} 193.87 $\angstrom$~intensity maps, Ar {\sc xiv}/Ca {\sc xiv} ratio maps without and with \emph{SDO}/HMI contours of $\pm$500 (white/magenta for positive/negative polarities) before (top section) and during the decay phase of the X9.3 flare (bottom section) on 2017 September 6.   The color bar scale shows the FIP effect as blue/green, photospheric composition as orange, I-FIP effect as yellow.  All \emph{Hinode}/EIS maps are co-aligned to \emph{SDO}/HMI maps at the times shown.  (P$_{0}$ is located at X = 492$\arcsec$, Y = -248$\arcsec$ at 03:23 UT and X = 590$\arcsec$, Y = -242$\arcsec$ at 16:43 UT). \label{fig_study559}}
\end{figure*}
%======================Figure 3 ============================

%======================Figure 4 ============================
\begin{figure*}[htp]
\centering
\includegraphics[width=0.75\textwidth]{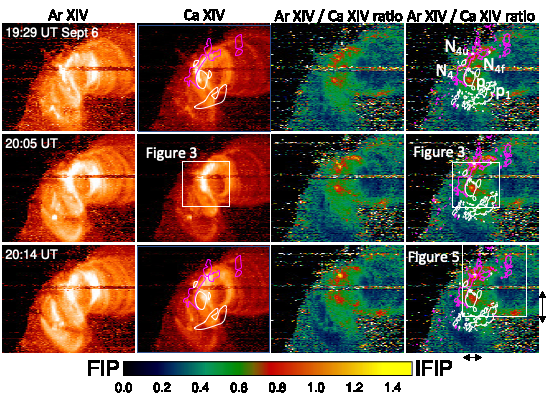}
\caption{Left to right: \emph{Hinode}/EIS Ar {\sc xiv} 194.40 $\angstrom$~and Ca {\sc xiv} 193.87 $\angstrom$~intensity maps, Ar {\sc xiv}/Ca {\sc xiv} ratio maps without and with \emph{SDO}/HMI contours of $\pm$500 (white/magenta for positive/negative polarities).  Observations are from 19:29 UT to 20:14 UT on 2017 September 6.  The color bar scale shows the FIP effect as blue/green, photospheric composition as orange, I-FIP effect as yellow.  All \emph{Hinode}/EIS maps are co-aligned to \emph{SDO}/HMI maps at the times shown.  (P$_{0}$ is located at X = 614$\arcsec$, Y = -241$\arcsec$ at 20:14 UT).
\label{fig_eisobs1}}
\end{figure*}
%======================Figure 4 ============================

The studies listed in \tab{study} contain the high-FIP Ar {\sc xiv} 194.40 $\angstrom$ (FIP = 15.76 eV) and low-FIP Ca {\sc xiv} 193.87 $\angstrom$ (FIP = 6.11 eV) emission lines.
The two ions have similar contribution functions therefore their intensity ratio is suitable to determine (I)FIP levels within the AR \citep{feldman09}.
In line with \citet{doschek15}, \citet{doschek16,doschek17} and \citet{baker19}, we use the following log$_{10}$ abundance values relative to log$_{10}$H = 12: 
Ca = 6.93 \citep{feldman92} and 6.33 \citep{caffau11} for the corona and photosphere, respectively; Ar = 6.5 \citep{lodders08} is the same for the corona and the photosphere.
Calculations of the contribution functions using these abundances were performed with the software included in the CHIANTI Atomic Database, version 8.0 \citep{dere97,delzanna15}.
Based on the intensity ratio of the contribution functions of high-FIP Ar {\sc xiv} and low-FIP Ca {\sc xiv}, values $>$1 indicate the I-FIP effect, = 1 is photospheric plasma, and $<$1 is FIP effect or typical solar coronal plasma.
The estimated uncertainty of the ratio is $\pm$0.28 assuming an intensity error of 20$\%$.
Considering the uncertainties, in the text and figures we refer to ratio values $\geq$ 1.3 as I-FIP effect plasma. 

Though Ar {\sc xiv} and Ca {\sc xiv} have similar contribution functions, it is not immediately apparent what, if any, temperature effects flaring might have on the uncertainties of the I-FIP effect measurements as large flares can have much higher densities (N) at high temperatures (T) \citep[e.g.][]{watanabe10,graham11,simoes15} compared to non-flaring active regions \citep{doschek07,tripathi08}.
The Ar {\sc xiv}/Ca {\sc xiv} ratio can vary by a factor $\sim$4 in the log$_{10}$ T = 6.6--6.9 range \citep[with T in K, see e.g.][]{doschek17,baker19}.

We derived the mean temperatures and densities for all strong I-FIP effect regions corresponding to Table \ref{tab_regions} where the appropriate diagnostic line pairs were included in the EIS studies (e.g. Ar {\sc xiv} 187.9/194.4 $\angstrom$ and Fe {\sc xiii} 203.8/202.0 $\angstrom$).
For raster times from 19:29--20:14 UT, we were only able to obtain temperature measurements.
Log$_{10}$ N were in the range 10.2--11.1 (with N in cm$^{-3}$) and the log$_{10}$ T were 6.6--6.7.
These compare to values in FIP effect regions of log$_{10}$ N = 8.9--9.9 and log$_{10}$ T = 6.2--6.4, respectively.

Higher density in the I-FIP effect regions lowers the theoretical ratio, so high observed I-FIP effect values of Ar {\sc xiv}/Ca {\sc xiv} become more extreme in the abundance differences that they imply.
Furthermore, the temperatures fall within the range that is considered to be plausible for the formation temperatures of Ar {\sc xiv} and Ca {\sc xiv} \citep{doschek17}.
It is very difficult to disentangle the density and temperature effects caused by the flaring but the increase in density combined with the fact that the temperature does not exceed log$_{10}$ T = 6.7 suggests that the density and temperature effects are unlikely to substantially affect the estimated uncertainty of $\pm$0.28. 
The results may be very different for measurements made during the peak phase of large flares but this is not the case here as the rasters were timed during the decay phases of the flares on September 6.

Spectroscopic data were reduced using standard routines available in the \emph{Hinode}/EIS section of Solar Software \citep[SSW;][]{freeland98}.
Three Gaussians were fit to the Ar {\sc xiv} 194.40 $\angstrom$ emission line to remove two unidentified lines in its blue wing. The Ca {\sc xiv} 193.87 $\angstrom$ line was also fit with three Gaussians to separate the line from nearby Fe {\sc x} 193.72 $\angstrom$, Ni {\sc xvi} 194.05 $\angstrom$, and 194.10 $\angstrom$ lines \citep{brown08,doschek15,baker19}.

All \emph{Hinode}/EIS maps and \emph{SDO}/HMI line-of-sight magnetograms were co-registered in a two step process.
First,  the aia$\_$prep.pro routine located in the SDO branch of SSW was used to align HMI and AIA data.
The routine takes into account differences in the plate scales and roll angles between the two instruments.
Second, either the Fe {\sc xii} 195.12 $\angstrom$ or 186.88 $\angstrom$ EIS intensity maps were aligned by eye with the co-temporal AIA 193 $\angstrom$ passband images.
The EIS Fe {\sc xii} emission lines and AIA 193 $\angstrom$ broadband imager sample plasma at similar temperatures  so that the same coronal structures are identifiable in each image making alignment straightforward between \emph{Hinode}/EIS and \emph{SDO}.
To help the reader to track different features in the figures, we have added arrows to indicate the central positions of major polarities and the distances between them.
A vertical arrow shows the positions and distance of the P$_{0}$--P$_{4}$ centers ($\sim$50$\arcsec$) and a horizontal arrow shows the same for P$_{0}$--N$_{4}$ ($\sim$30$\arcsec$).  

\fig{goes}, lower panel, shows a sample Ar {\sc xiv}/Ca {\sc xiv} ratio map with/without \emph{SDO}/HMI contours of $\pm$500 G (white/magenta for positive/negative polarities) at 19:56 UT on September 6.
Four regions of interest based on the magnetic flux systems described in \sect{b_obs} are labeled as follows:  n$_{2}$ in the spatially extended magnetic field to the north; N$_{4u}$ which detached northward from N$_{4}$; p$_{2}$ located on the east side of the large positive sunspot P$_{0}$; and p$_{1}$ along the extended positive polarity to the south.
Similar ratio maps with the corresponding Ar {\sc xiv} 194.40 $\angstrom$ and Ca {\sc xiv} 193.87 $\angstrom$ intensity maps are displayed in \figs{study559}{eisobs1}.

\emph{Hinode}/EIS observed AR 12673 at 02:35 UT on September 6 during a relatively quiet period (\fig{study559}, top panel).
The core of the active region is predominantly composed of typical coronal plasma.
FIP effect levels range from $\sim$0.3 to 0.5 using the high-FIP Ar {\sc xiv}/low-FIP Ca {\sc xiv} ratio, which is equivalent to the conventional solar low-FIP/high-FIP ratio of 2.0--3.3.
Localized regions at p$_{2}$ and N$_{4}$ show weak indications of I-FIP effect plasma. 
Both patches of I-FIP effect plasma evolve to photospheric composition within $\sim$25 min (at 02:59 UT) and the patch associated with N$_{4}$ returns to coronal plasma within $\sim$50 min (at 03:23 UT) when \emph{Hinode}/EIS composition observations ended.

EIS composition observations resumed at 16:13 UT during the extended decay phase of the X9.3 flare, just after the peak of an M2.5 flare at 15:56 UT (see the \emph{GOES} soft X-ray light curve in \fig{goes}).  
At this time plasma of photospheric composition was present in the southern part of  n$_{2}$ and at N$_{4u}$.  
Thirty minutes later, the plasma at N$_{4u}$ had evolved to I-FIP effect plasma while weak I-FIP patches appeared at p$_{2}$ (\fig{study559}, bottom panel).
I-FIP effect  plasma was observed at these locations until the \emph{Hinode}/EIS observing sequence ended at 17:07 UT

%======================Table 2 =========================
\begin{table}[h]
	\centering
	\caption{Mean I-FIP effect ratio values on September 6 in regions defined in \fig{goes} with the selection of pixels having ratio values $\geq$ 1.3.}
	\label{tab_regions}
	\begin{tabular}{ccccc}
		\hline
		  Raster Times (UT) & n$_{2}$  & N$_{4u}$ & p$_{2}$ & p$_{}$ \\
		\hline
		04:11 & -- & 1.5 & -- & --\\
		16:19 & 1.5 & 1.4 & -- & --\\
		16:25 & -- & 1.4 & -- & --\\
		16:31 & -- & 1.5 & -- & --\\
		16:37 & -- & 1.4 & -- & --\\
		16:43 & -- & 1.3 & -- & 1.4\\
		16:49 & -- & 1.3 & -- & 1.4\\
		16:55 & -- & 1.3 & -- & 1.3\\
		17:01 & 1.3 & 1.4 & -- & --\\
		19:29 & 1.8 & -- & -- & 1.3\\
		19:38 & 1.3 & -- & 1.5 & 1.8\\
		19:47 & 1.4 & -- & 1.3 & 1.3\\
		19:56 & 1.6 & 1.4 & 1.3 & 1.4\\
		20:05 & 1.4 & 1.4 & -- & 1.5\\
		20:14 & 1.4 & 1.3 & -- & 1.5\\
		\hline
			\end{tabular}
\end{table}
%======================Table 2 ============================

The data shown in \fig{eisobs1} were obtained when
\emph{Hinode}/EIS was operating in its autonomous observing mode.
A flare response study was triggered by an M1.4 flare at 19:29 UT on September 6.
The large FOV is centered on the active region and extends to the flare loops which is not the case in \fig{study559} where the smaller FOV covers only the core of the active region (see the FOVs shown in white in \fig{eisobs1}). 
All regions defined in \fig{goes} have patches of I-FIP effect plasma from 19:29--20:14 UT except for p$_{2}$ where the plasma showing weak indications of the I-FIP effect evolved to photospheric composition from 20:05 to 20:14 UT.

The mean I-FIP effect ratios for pixels with values $\geq$ 1.3 for each of the regions defined in the lower left panel of \fig{goes} are given for \emph{Hinode}/EIS rasters on September 6 in \tab{regions}.
In general, I-FIP effect plasma persisted in the n$_{2}$ and p$_{1}$ regions and at N$_{4u}$, but less so within p$_{2}$.

%======================Figure 5 ============================
\begin{figure*}[t]
\centering
\includegraphics[width=0.72\textwidth]{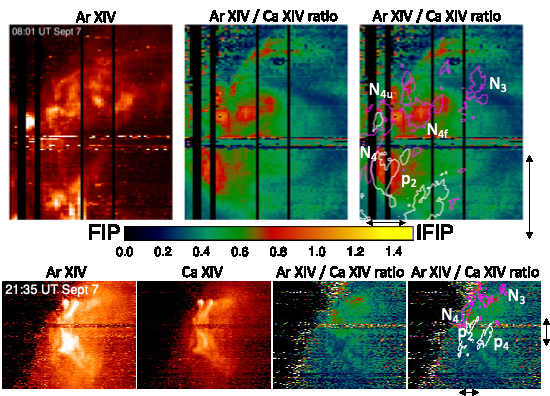}
\caption{Top panel:  \emph{Hinode}/EIS Ar {\sc xiv} intensity and Ar {\sc xiv}/Ca {\sc xiv} ratio maps without and with \emph{SDO}/HMI contours of $\pm$500 (white/magenta for positive/negative polarities) on 2017 September 7 at 08:01 UT.  Bottom panel:  \emph{Hinode}/EIS Ar {\sc xiv} 194.40 $\angstrom$~and Ca {\sc xiv} 193.87 $\angstrom$~intensity maps, Ar {\sc xiv}/Ca {\sc xiv} ratio maps without and with \emph{SDO}/HMI contours on  2017 September 7 at 21:35 UT.  (P$_{0}$ is located at X = 692$\arcsec$, Y = -232$\arcsec$ at 08:01 UT and at X = 766$\arcsec$, Y = -222$\arcsec$ at 21:35 UT).
\label{fig_eis_sep7}}
\end{figure*}
%======================Figure 5 ============================
 
On September 7 at 08:01 UT, ~photospheric  plasma was present within the flare loops associated with p$_{2}$ and at both pieces of the well-separated breakaway negative polarity N$_{4u}$ and N$_{4f}$ in the top panel of \fig{eis_sep7}.
This single \emph{Hinode}/EIS observation was acquired during the decay phase of a C8.2 flare that peaked at 06:42 UT.
The plasma within the p$_{1}$ region had returned to FIP effect composition in the Ar XIV/Ca XIV ratio maps.
Approximately 13.5 hr later, a C5.4 flare triggered a series of \emph{Hinode}/EIS observations from 21:35 to 22:19 UT. 
\fig{eis_sep7}, bottom panel, shows a representative sample at 21:35 UT.
I-FIP effect plasma was no longer visible within the active region.
Only remnant patches of photospheric composition remained in the vicinity of  N$_{4u}$.
Coronal composition was observed everywhere else including at p$_{2}$ where patches of I-FIP effect plasma were present on the previous day.

%%%%%%%%%%%%%%%%%%%%%%%%%%%%%%%%%%%%%%%%%%%%%%%%%%%%%%%%%%%%%%%%%%%%%%%%%%%%%%%%%%
\section{Discussion and Interpretation} \label{sect_interp}
In this study, we have analyzed the evolution of plasma composition in AR 12673 on 2017 September 5--7. 
\emph{Hinode}/EIS detected highly localized patches of I-FIP effect plasma embedded within the characteristic FIP effect plasma of the active region core \citep[e.g.][]{gdz14,baker15}.
I-FIP effect  plasma was present $\sim$8 hours before the X9.3 class flare at 12:02 UT on September 6 and also during the long decay phase extending to $\sim$20:00 UT. 
The plasma composition of AR 12673 had returned to FIP effect when \emph{Hinode}/EIS observed it $\sim$13 hr later.

%======================Figure 6 ============================
\begin{figure*}[htp]
\centering
\includegraphics[width=0.7\textwidth]{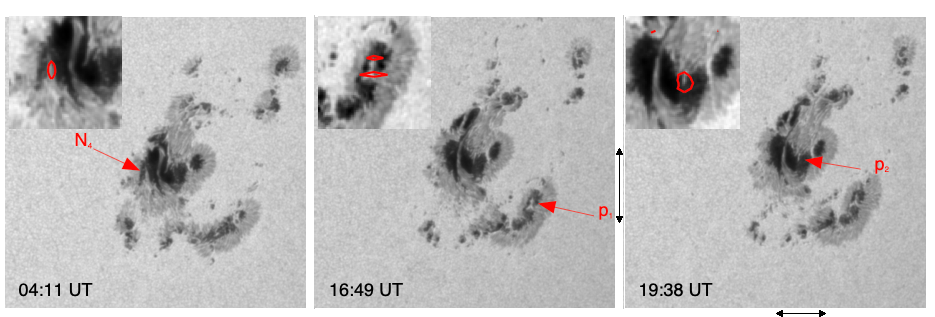}
\includegraphics[width=0.71\textwidth]{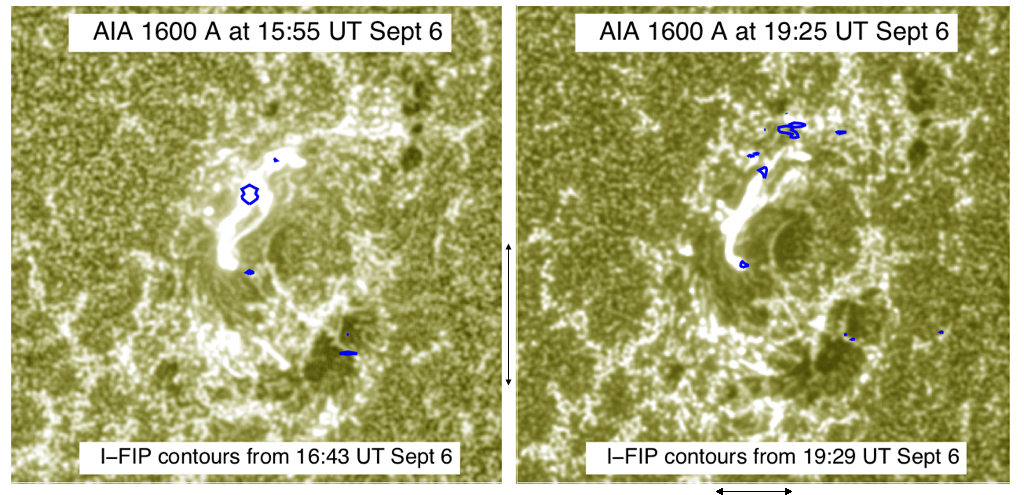}

\caption{Top panel (left to right):  \emph{SDO}/HMI continuum images at 04:11, 16:49, and 19:38 UT on 2017 September 6.  Red arrows indicate locations of strong light bridges at N$_{4}$, p$_{1}$, and p$_{2}$ regions defined in Figure \ref{fig_goes}.  Insets are zoomed images of the light bridges overplotted with red contours of I-FIP effect $=$ 1.3. 
The I-FIP effect plasma is precisely at the locations of the strong light bridges.  Bottom panel:   \emph{SDO}/AIA 1600 $\angstrom$ images overplotted with blue I-FIP effect plasma contours of $\geq$ 1.3.  Contours have been derotated from the times of the \emph{Hinode}/EIS rasters.  (P$_{0}$ is located at X = 610$\arcsec$, Y = -241$\arcsec$ at 19:38 UT and at X = 608$\arcsec$, Y = -241$\arcsec$ at 19:25 UT).
\label{fig_aia_cont}}
\end{figure*}
%======================Figure 6 ============================

%======================Figure  7 ============================
\begin{figure*}[t]
\centering
\includegraphics[width=0.62\textwidth]{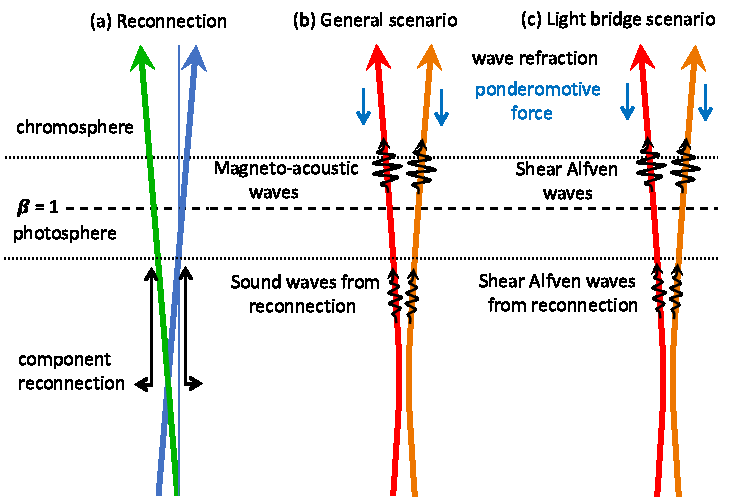}
\caption{(a) Two flux systems with a non-zero component of anti-parallel field between them are pushed together and reconnect below the photosphere.  In the case of the general scenario shown in (b), reconnection generates sound waves that mode convert into magneto-acoustic waves at the $\beta$ = 1 layer.  These waves are refracted/reflected in the chromosphere, leading to a downward oriented ponderomotive force which depletes the low-FIP ions from the chromosphere.  The general scenario is consistent with \cite{baker19}.  In the specific light bridge scenario (c) proposed in Section \ref{sec_pfmodel}, sub-photospheric reconnection launches incompressible (shear) Alfv\'en waves along the magnetic field lines at the edges of the light bridge.  No mode conversion takes place but the rest of the scenario in (b) remains.  \label{fig_scenario}}
\end{figure*}
%======================Figure 7 ============================
 
 \subsection{I-FIP Effect Plasma Observed at the Locations of Sub-photospheric Reconnection}
\label{sect_disc_reconnect}
 A notable feature of the evolution of the highly complex magnetic field of AR 12673 is the coalescence of same polarity magnetic field strands in a number of locations throughout the active region.
 From the time of its emergence on September 3, the positive polarity strands of the E--W oriented arcade system merged together east of P$_{0}$ until a coherent umbra was evident at p$_{2}$ in the continuum images early on September 4 (see Fig1$\_$movie.mp4).
 A train of positive umbrae approaching from the north continued to coalesce with the already coherent p$_{2}$ umbra for the next $\sim$2 days before the structure began to fragment on September 7.
 
The main N--S oriented flux system emerged after the arcade system later on September 3 when the opposite polarities rapidly separated, N$_{4}$ moving in a northward direction and P$_{4}$ progressing to the south-southwest.
Right from the start, there was a continuous influx of new strands of negative polarity coalescing at N$_{4}$ for the same time period as with p$_{2}$.
Co-temporal emergence of P$_{4}$ with N$_{4}$ forced positive field in the vicinity of p$_{1}$ to amalgamate into the sunspot in the southwest of the active region.
The process of sunspot coalescence driven by strong flux emergence of same polarity field was repeated in n$_{2}$ when N$_{4u}$ broke away from the main N$_{4}$ umbra.
It pushed into the negative field at n$_{2}$ and then rotated around P$_{3}$.

I-FIP effect plasma was observed within the coalescing umbrae at p$_{2}$, N$_{4}$, P$_{4}$--p$_{1}$, and N$_{4u}$--n$_{2}$.
The patches can be located more precisely within the umbrae where strong LBs formed during the merging and coalescence of the sunspots.
So-called strong LBs appear between magnetic field strands of the same polarity during the early stages of the AR's development rather than during sunspot decay which is more commonly associated with faint LBs \citep[e.g.][and references therein]{felipe16}.
Red arrows indicate the locations of strong LBs within N$_{4}$, p$_{2}$, and p$_{1}$ in the HMI continuum images from September 6 in Figure \ref{fig_aia_cont} (top panels).
Contours of I-FIP effect plasma are overplotted on zoomed images of the LBs.
The I-FIP effect plasma is precisely located at the LBs within the coalescing umbrae of the respective regions.

Once major flux emergence paused in the core of the AR late on September 6, the LBs were either no longer visible (N$_{4}$) or they became part of the fragmentation process in the sunspot during its decay phase (p$_{2}$).
On the southern edge of the AR, the LB at p$_{1}$ remained prominent for at least another day as flux emergence was still ongoing, driving P$_{4}$ into p$_{1}$ and forcing coalescence of the sunspot.
The evolution of the LBs can be viewed in the HMI continuum images of Fig1$\_$movie.mp4.

Coalescing sunspot umbrae are preferential sites for sub-chromospheric reconnection to take place especially during the early development of highly complex active regions.
Major flux emergence episodes push flux tubes of the same polarity against each others.
From their different sub-photospheric evolution, these flux tubes are expected to typically have a finite angle between them.  Then, component magnetic reconnection is expected to be present between the coalescing flux tubes (see Figure \ref{fig_scenario}a). 

There is a plausible alternative scenario of internal reconnection within an individual flux tube in the case of N$_{4u}$.
The flux tube may have been deformed as it rotated around P$_{3}$, creating magnetic shear, beginning at around the time of the X2.2 flare and continuing during the early decay phase of the X9.3 flare.
This evolution is shown from early on September 6 in the movie linked to Figure \ref{fig_br_map}.
Significant deformation of the N$_{4u}$ flux tube may generate internal current sheets where sub-photospheric reconnection could take place.
Either scenario or a combination of them is plausible at N$_{4u}$, however internal reconnection does not appear to be a likely cause of reconnection at p$_{2}$, N$_{4}$, and p$_{1}$ which showed little evidence of magnetic shear in their formation.  

%======================Table 3============================
\begin{table}[h]  
	\centering
	\caption{Flare class, flare start, peak, and end times from 2017 September 6 with the locations of flare ribbons corresponding to Figures \ref{fig_br_map} and \ref{fig_goes}.  Flare details are from LMSAL SolarSoft Latest Events: \url{http://www.lmsal.com/solarsoft/latest_events_archive/}. $^*$ The X9.3 flare loops are still present well after this time so that the decay phase of the X9.3 flare is mixed with the following flares. This is also the case for other M and X flares. 	}
\label{tab_flares}	
	\begin{tabular}{ccc} 
		\hline
		Flare Class & Start/Peak/End & Flare \\
		& Times (UT) &  Ribbons \\
		\hline		
C2.7 & 07:29/07:34/07:48        & p$_{2}$,\, N$_{3}$--N$_{4}$\\
X2.2 & 08:57/09:10/$>$11:53     & p$_{2}$,\, N$_{3}$--N$_{4}$\\
X9.3 & 11:53/12:02/$>$15:51$^*$ & p$_{2}$--P$_{3}$--P$_{4}$,\, N$_{4u}$--N$_{4}$ \\
M2.5 & 15:51/15:56/$>$19:21     & p$_{2}$--P$_{3}$,\, N$_{3}$--N$_{4u}$--N$_{4f}$ \\
M1.4 & 19:21/19:30/$>$23:33     & p$_{2}$--P$_{3}$,\, N$_{3}$--N$_{4u}$--N$_{4f}$ \\
M1.2 & 23:33/23:39/23:44        & p$_{2}$,\, N$_{4u}$--N$_{4f}$\\
		\hline
\end{tabular}
\end{table}
%======================Table 3============================

\subsection{Revealing I-FIP effect  plasma with Flare Energy Input}

Successive episodes of major flux emergence and its interaction with a pre-existing, well-anchored sunspot resulted in the formation of the highly complex magnetic structure of AR 12673.
Continual interaction gave rise to significant flare productivity \citep[e.g.][]{yang17,verma18,romano18,romano19,toriumi19}, indicating the occurrence of magnetic reconnection in the solar corona.

 Evidence of significant localized heating generated by flaring is provided in \tab{flares} which lists the details for all flares $>$ C1.0 on September 6:  flare class, timings (start/peak/end), and locations of flare ribbons.
The locations of the flare ribbons were identified either explicitly as ribbons in the \emph{SDO}/Atmospheric Imaging Assembly (AIA) \citep{lemen12} 1600 $\angstrom$ or in AIA 193 $\angstrom$ images (not shown) as post-flare loops, the coronal counterpart of the ribbons.
Major flaring occurred in the vicinity of the U-loop configuration introduced in Section \ref{sect_b_obs} \citep{yang17,yan18}.
 
 During the rise phase of the X9.3 flare, two ribbons began to develop on either side of the C-shaped PIL within the active region core involving mainly P$_{3}$--p${_2}$--P$_{4}$ and N$_{4u}$--N$_{4}$.  
The ribbons then extended into N$_{4f}$ and N${_3}$ soon after the eruption of the magnetic flux rope/CME coinciding with the flare peak at 12:02 UT \citep{mitra18}.
The overlying post-flare loop arcade rooted in the two flare ribbons remained visible in the \emph{SDO}/AIA coronal passbands throughout the extended decay phase of the flare until the early hours of September 7.

Ribbons associated with the M2.5 and M1.4 flares (peaking at 15:56 and 19:30 UT, respectively) also appeared along the PIL, tracing parts of the X9.3 ribbon `tracks' at p$_{2}$--P$_{3}$ and N$_{3}$--N$_{4u}$--N$_{4f}$.
The ribbons are clearly evident in \emph{SDO}/AIA 1600 $\angstrom$ images at 15:55 and 19:25 UT in the bottom panel of Figure \ref{fig_aia_cont}.
Blue contours of I-FIP effect plasma (ratio $\geq$ 1.3) are overplotted on the images and have been derotated from the times of the \emph{Hinode}/EIS raster times given in the figure.
Contours located at p$_{2}$ and N$_{4u}$ lie directly along the ribbons of both flares as does the n$_{2}$ contour associated with the M2.5 flare.
There are no obvious flare ribbons extending to P$_{4}$ in the AIA 1600 $\angstrom$ image at 15:55 UT, however, the I-FIP effect contours correspond to the footpoints of the post-flare arcade at 16:43 UT (not shown).

\subsection{Comparison of ARs 11429 and 12673}
The \emph{Hinode}/EIS observations of plasma composition in AR 12673 are globally comparable to those of AR 11429 \citep{baker19}.
Distinct patches of I-FIP effect  plasma were located at coalescing sunspot umbrae during flaring activity.
The patches were observed when and where flare ribbons crossed the umbrae.

In the case of AR 11429, two compact patches of I-FIP effect plasma appeared and disappeared within one hour during the decay phase of a single, isolated M-class flare.
Though the scenarios are very similar, the scales are different for AR 12673 in that four patches were present for time scales of hours not minutes over the merging umbrae of four instead of two sunspots.
Furthermore, soft X-ray emission was significantly elevated for hours leading up to the first X-class flare and continued at very high levels for over 16 hrs after the X9.3 flare peak (Figure \ref{fig_goes}) compared to less than two hours during the M-class flare in AR 11429.
The two M-class flares that occurred during the extended decay phase of the larger of the two X flares added additional heating sources. 
This does not necessarily mean that an increase in flaring activity directly produces more I-FIP effect plasma since more flaring would only allow us to observe more I-FIP effect plasma regions via chromospheric evaporation if they are present. 

 However, more globally, an increase of magnetic complexity can induce more reconnection below the fractionation region (in the chromosphere) and in the corona implying both the creation of more I-FIP effect  plasma and more flares. Then, we expect a generic correlation between the complexity and emergence rate of the magnetic field, the amount of I-FIP effect  plasma, and the flaring activity level.

Another common feature to both active regions is the occurrence of patches of I-FIP effect plasma at LBs within the coalescing umbrae.
After noticing the precise location of the patches at the strong LBs during the formation stages of the same polarity sunspot umbrae, we re-examined the continuum movie of AR 11429 and indeed, patches of I-FIP effect plasma were also found under similar circumstances at strong LBs.

\subsection{Why is I-FIP effect plasma observed at light bridges?}\label{sec_pfmodel}
\citet{baker19} proposed that transient patches of I-FIP effect plasma observed in AR 11429 were created by increased fast mode wave flux 
that was generated by sub-chromospheric/-photospheric reconnection of coalescing umbrae (Figure \ref{fig_scenario}b).
Fast mode waves coming from below the fractionation region of the chromosphere and undergoing a total internal reflection mean that the ponderomotive force is directed downward so that low-FIP elements are depleted from the chromospheric plasma \citep{laming15}.
Sunspot umbrae are preferential locations to observe I-FIP effect  plasma.
 They are where the magnetic field is strongest therefore the plasma $\beta$ = 1 layer is lower in the photosphere
 \citep{avrett15} so that the fast mode wave flux from below is enhanced.
 The vertical field of the umbrae is also where the downward-directed ponderomotive acceleration is likely to be the strongest as the plasma upflow must be along the magnetic field, making the field-aligned ponderomotive force most relevant.  

EIS observed I-FIP effect plasma patches in four regions where sub-chromospheric or even sub-photospheric reconnection was likely to be an ongoing process.
In all cases, significant flux emergence was the main driver of sunspot coalescence.  
This is consistent with the findings of \citet{baker19}.
The classical scenario is forced magnetic reconnection between two independent flux tubes where there is some angle between the same polarity strands that are pushed together by convective motions (Figure \ref{fig_scenario}a), leading to the formation of MHD waves in the process. 
AR 12673 is an extreme case of the forced reconnection scenario due to the amount and complexity of the emerging magnetic field. 
There is essentially a rescaling due to energy input.
Finally, I-FIP effect  plasma was not observed within the AR after major flux emergence of the main flux system ceased. 

Previously, we have argued in \cite{baker19} that fast mode waves cause the I-FIP effect, mainly because of the ease, relative to Alfv\'en waves, of achieving the required degree of reflection to produce I-FIP rather than FIP effect fractionation.
Fast mode waves (sound waves) produced below the $\beta=1$ layer by reconnection mode convert where $\beta=1$ to continue as fast modes (now magneto-acoustic waves) in the $\beta<1$ chromosphere, where they refract/reflect exerting a downward ponderomotive force on the ions (Figure \ref{fig_scenario}b).
The mode conversion is essential, and increases with increasing angle between the magnetic field direction and the wave vector.
While this might produce some selectivity in where I-FIP can be produced, upward propagating sound waves are less likely to mode convert and produce I-FIP in the vertical magnetic field, and it is still less obvious why I-FIP should be restricted to the LBs, as observed in AR 11429 and AR 12673.

We propose in the following a slightly more refined scenario, which consists of two interlinked parts: the first ingredient is the geometry of the sub-photospheric reconnection that creates the train of waves responsible for the I-FIP effect fractionation; the second one is the nature of waves such that they remain associated with a specific bundle of field lines, emerging at a very specific location, and only there. 

LBs are very particular areas within an active region where pockets of relatively unmagnetized plasma are trapped within two merging flux systems of the same strong polarity, forming a sunspot. 
The sub-photospheric forces that merge the two flux systems may eventually be able to squeeze out the unmagnetized plasma and lead to a structure that looks like a single-polarity spot in white light images and line of sight magnetograms.
Therefore, LBs are the photospheric trace of the merging plane between two flux systems that, generally, extend down in the convection zone, at least deeper than the vertical extension of the unmagnetized plasma volume that supports the LB. 
At the sub-photospheric location where the two flux systems are pushed together by the high-$\beta$ plasma it is conceivable that component field reconnection between nearly parallel flux bundles of the same sign occurs (Figure \ref{fig_scenario}a). 
This is the main assumption of this scenario.
If this is true, then the numerical simulations by \cite{kigure10} provide crucial information, namely that in a high-$\beta$ plasma environment, the reconnection between nearly-parallel flux systems would mostly generate incompressible (shear) Alfv\'en waves.   
Applying this result in our hypothetical scenario (Figure \ref{fig_scenario}c), the sub-photospheric reconnection process would then launch incompressible Alfv\'en waves along the field lines of the flanks of the reconnecting flux systems that surrounds the LB.
Since Alfv\'en waves must propagate along field lines, as they travel upward, these waves would then eventually emerge at the sides of the LB, irrespective of how deep the LB extends below the photosphere.   
Furthermore, the magnetic field right above a LB was observed recently by \cite{felipe16} using high resolution measurements from the GREGOR telescope \citep{schmidt12}. 
The authors conclude that the field at the side of the LB stretches up along its two sides and converges above the LB,  forming a cusp-shape across it, with the mainly horizontal LB field being confined below the cusp.  
Following the cusp lines, the shear Alfv\'en waves of sub-photospheric origin would reach the chromosphere, where they could provide the energy, and the correct direction of travel, for the ponderomotive I-FIP effect fractionation in a very localized volume, namely right above the LB. 
Hence, if the above scenario is correct,  the observations of I-FIP effect fractionated plasma should be located between the merging sunspot's fragments, i.e. at the LB, which is indeed what is observed. 

The above scenario is highly speculative.
Getting sufficient reflection of Alfv\'en waves to cause I-FIP rather than FIP effect fractionation probably requires more wave interaction physics in the chromospheric model, as Alfv\'en waves likely reflect off sound waves, shocks or other density inhomogeneities.
This can be modeled analytically, but will require both numerical testing and observational confirmation.
The numerical testing is needed to confirm that upwards traveling shear Alfv\'en waves from below the photosphere can reflect efficiently and drive the I-FIP effect fractionation.
The observational verification entails a multi-line spectropolarimetric study of LBs to verify the presence of such waves at the sides of the LB. 
On the other hand, such a scenario assumes sub-photospheric reconnection between the merging flux systems that drives the formation of an observed LB  and therefore determines the observed localization of I-FIP effect plasma on the basis of the properties of reconnection between nearly-parallel flux systems. 
As mentioned above, such a localization at LBs is more difficult to explain with compressible waves which can travel across field lines. 
%%%%%%%%%%%%%%%%%%%%%%%%%%%%%%%%%%%%%%%%%%%%%%%%%%%%%%%%%%%%%%%%%%%%%%%%%%%%%%%%%%
\section{Conclusions}\label{sect_conc}
I-FIP effect plasma has been detected in only eight active regions \citep{doschek15,doschek16,doschek17}, but it is not yet clear how widespread the phenomenon is. 
Observational constraints are likely to play a role in limiting the number of I-FIP effect plasma detections. 
It requires the right composition lines in the employed study, the correct target selection, and pointing at the specific flaring locations within the target active region during the time of flaring activity. 

A significant fraction of the I-FIP effect plasma observations were obtained while \emph{Hinode}/EIS was operating in its flare trigger mode. This is an observing strategy to catch flares, and the response study is initiated only at or near
the flare peak time. It remains to be seen whether a different observing strategy, dedicated to observing I-FIP effect events, would be more successful at capturing the evolution of plasma composition before and during the flare rise phase.

Notwithstanding these limitations, observations of AR 11429 and AR 12673 show that they share common characteristics in their magnetic configuration which distinguish them as potential candidates for the detection of I-FIP effect plasma.
Both active regions were large ($>$ 500 MSH) and magnetically complex ($\beta \gamma \delta$) with ongoing major flux emergence of sheared/twisted field which violently interacted with pre-existing field.

Our interpretation of the \emph{Hinode}/EIS observations of I-FIP effect  plasma in AR 12673 is consistent with the ponderomotive fractionation model. 
Ongoing sub-chromospheric or sub-photospheric reconnection at multiple sunspot umbrae is likely to increase the wave flux from below the chromosphere thereby providing favorable circumstances for the accumulation of I-FIP effect  plasma in more locations and on longer time scales than for AR 11429.

Flaring and evaporation of chromospheric material are essential parts of this scenario.
Along flare ribbons at the footpoints of newly formed loops, which rarely traverse coalescing umbrae, the expected chromospheric composition is close to photospheric.
The Sun-as-a-star \citep{warren14} and stellar \citep{audard03,g-a09,testa15} observations during flares are consistent with this since they indicate that the overall elemental composition is getting close to photospheric value.
The presence of bright I-FIP patches shifts the overall composition of FIP-effect dominated coronae even more towards photospheric or indeed to I-FIP effect values \citep[e.g.][]{katsuda20}. 

Do we need `monster' active regions to produce I-FIP effect coronal composition? 
So far the evidence is pointing that way, but the picture is clouded by observational constraints. 
What is likely to be a requirement is a strong wave flux from below the chromospheric fractionation region. 
There is a considerable fraction, $\sim$30$\%$, of the total released energy of Alfv\'en waves generated in high plasma--$\beta$ reconnection events \citep{kigure10}, so that I-FIP effect is expected to occur more often than what has been observed so far if sub-photospheric reconnection occurs frequently between the multiple thin flux tubes present in emerging ARs.
Based on the observations of AR 12673, we propose a scenario where shear Alfv\'en waves originating below the photosphere and traveling upward can localize the I-FIP effect plasma at strong light bridges at coalescing umbrae.

Active region complexity is highly correlated with activity \citep[see e.g.][and references therein]{toriumi19} so there is a temptation to associate the presence of localized regions of I-FIP effect  plasma in so-called monster active regions on the Sun with more active M-dwarf stars containing very large/strong starspots \citep[e.g.][]{berdyugina05,reiners12} and whose coronae are dominated by I-FIP effect  plasma.
It is expected to be a matter of the scale of waves generated by magnetic reconnection in sub-fractionation layers of stars.
X-ray flux is a function of overall heating in a star's atmosphere \citep[e.g.][]{wood18} which in turn governs the elements that are ionized on that star.
 The presence of strong and highly complex magnetic field provides additional heating and MHD waves that determine the direction and strength of the ponderomotive force and therefore the degree and direction of plasma fractionation, whether I-FIP or FIP effect, on stars of spectral types F--M.

The filling factor of spot umbrae on the photosphere of a star is likely to be a significant factor in determining the extent to which I-FIP effect  plasma fills the star's corona.  
On the Sun, strong magnetic fields are found in the small localized areas of sunspots covering a small fraction of the surface \citep[$<$ 0.5$\%$,][]{hathaway15} hence we may observe very small patches of I-FIP effect plasma.
On the more active M-dwarfs, starspots can cover up to 2 orders of magnitude more of a star's photosphere \citep[e.g.][]{jackson13,tregloan19}, which may explain why their coronae are dominated by I-FIP effect plasma.

The recognition that stellar coronal composition has a dependence on magnetic activity  \citep[][]{audard03,g-a09,testa15} is in line with our results of resolved I-FIP generation on the Sun in a FIP-effect dominated low-activity star.
The I-FIP patches are closely related to strong magnetic field concentrations and made observable in the corona by flaring.

\acknowledgments
We thank the referee for their helpful suggestions to improve the clarity of the manuscript.
The authors are very grateful to Konkoly Observatory, Budapest, Hungary, for hosting two workshops on Elemental Composition in Solar and Stellar Atmospheres (IFIPWS-1, 13-15 Feb, 2017 and IFIPWS-2, 27 Feb-1 Mar, 2018) and acknowledge the financial support from the Hungarian Academy of Sciences under grant NKSZ 2018$\_$2. 
The workshops have fostered collaboration by exploiting synergies in solar and stellar magnetic activity studies and exchanging experience and knowledge in both research fields.
Hinode is a Japanese mission developed and launched by ISAS/JAXA, collaborating with NAOJ as a domestic partner, and NASA and STFC (UK) as international partners. 
Scientific operation of Hinode is performed by the Hinode science team organized at ISAS/JAXA. 
This team mainly consists of scientists from institutes in the partner countries. 
Support for the post-launch operation is provided by JAXA and NAOJ (Japan), STFC (UK), NASA, ESA, and NSC (Norway). 
\emph{SDO} data were obtained courtesy of NASA/\emph{SDO} and the AIA and HMI science teams.
D.B. is funded under STFC consolidated grant number ST/S000240/1 and L.v.D.G. is partially funded under the same grant.
The work of D.H.B. was performed under contract to the Naval Research Laboratory and was funded by the NASA Hinode program. 
GV acknowledges the support  from the European Union's Horizon 2020 research and innovation programme under grant agreement No 824135 and of the STFC grant number ST/T000317/1
J.M.L. was supported by the NASA HGI (80HQTR19T0029), NASA HSR (NNH16AC391), and LARS (NNH17AE601) programs, the Chandra GO program and by Basic Research Funds of the Chief of Naval Research.
DML is grateful to the Science Technology and Facilities Council for the award of an Ernest Rutherford Fellowship (ST/R003246/1).

%%BIBLIOGRAPHY%%%%%%%%%%%%%%%%%%%%%%%%%%%%%%%%%%%%%%%%%%%%%%%%%
\bibliographystyle{aasjournal}
\bibliography{uloop}

\end{document}